\documentclass[a4paper,11pt]{article}
\usepackage{jinstpub} 
\usepackage{lineno}
\usepackage{graphicx}
\usepackage{subcaption}
\usepackage{siunitx}
\usepackage{booktabs}
\title{\boldmath Analytical Modeling of Asynchronous Event-Driven Readout Architectures Using Queueing Theory}

\author[1]{Dominik S. Górni\note{Corresponding author.}}
\author{and Grzegorz W. Deptuch}
\affiliation{AGH University of Krakow, \\Faculty of Electrical Engineering, Automatics, Computer Science and Biomedical Engineering, 
\\Department of Metrology
and Electronics, 
\\al. A. Mickiewicza 30, 30-059 Krakow, Poland}

\emailAdd{dgorni@agh.edu.pl}

\abstract{
Event-driven imagers and sensor arrays commonly employ asynchronous arbiter trees with a synchronous acknowledge to serialize requests. We present an analytical framework that models the root as an \(M/D/1\) queue with deterministic quantum \(T\) and implements losses at the sources through one-slot gating. The admitted rate, loss probability, utilization, and mean sojourn time are coupled by self-consistent relations; a closed form for \(\mathbb{E}[S_t]\) separates fixed path delay \(\tau_0\) from queueing effects. The framework matches post-layout results of a physical prototype over light to heavy traffic, reproducing saturation at \(1/T\) and the observed latency growth, while classical \(M/G/1/K\) and Engset-type abstractions diverge at higher occupancy. Because all relations are algebraic, they enable rapid sizing at design time, including the impact of partitioning into independent tiles: reducing fan-in lowers arbitration depth and \(\tau_0\), decreases loss, and improves latency at fixed \(T\), with throughput adding across tiles. The model thereby links architectural parameters to performance metrics and supports selection of acknowledge period, tiling, and link count under practical constraints.
}


\begin{document}
\maketitle
\flushbottom

\section{Introduction}
Modern radiation detectors, particle trackers and intelligent sensor arrays increasingly rely on \emph{event-driven} readout to achieve high throughput at low power while preserving precise timing information~\cite{He2023, Schayck_2022}. In an event-driven system, pixels (or, more generally, any sensing channels) request access only when any activity occurs, and the readout architecture arbitrates among the outstanding requests without imposing a global frame~\cite{Gorni_prime}. This contrasts with frame-based schemes, in which the entire matrix is periodically sampled at a fixed rate. Even some pseudo-event-driven schemes that use only combinational logic for arbitration, like Address-Encoder Readout-Decoder (AERD) need to freeze the matrix state over coarse capture windows (typically \(2\text{–}10~\upmu\text{s}\)) to avoid dynamic switching~\cite{Yang_2019}. In a truly event-driven architecture, exemplified by EDWARD (Event-Driven With Access and Reset Decoder), a fully asynchronous binary-tree network of arbiters is used to grant bus access to requesting pixels, while a synchronous acknowledge clock provides the data-transfer quantum that allows full synchronization with the external data-acquisition system~\cite{Gorni_2022}. When a validated hit occurs, the corresponding pixel autonomously requests the shared bus -- if it wins arbitration, it transmits its address and optional payload, receives an acknowledge, and self-resets, readying the system for the next event.

An analytical description of such event-driven systems is essential at design time. Given project constraints, expected and peak per-channel rates, aggregate load, acceptable loss (pileup), and target timing resolution, an analytical model lets the designer quickly explore architectural trade-offs, including: (i) the number of independent arbitration trees (and their fan-in), (ii) the number of parallel outputs and serialization speed, (iii) the acknowledge period \(T\) and its distribution of delays along the tree, and (iv) buffering policies at the pixel and system levels. With a predictive model, one can select parameters that avoid saturation over the operating conditions while meeting timing-resolution requirements, rather than relying on time-consuming end-to-end simulations alone.

Timing resolution is a first-order benefit of true event-driven readout: every accepted hit can be time-stamped at the readout boundary. However, the achievable resolution is not solely set by the local time-stamp circuit -- it also depends on the \emph{service (sojourn) time} from event occurrence to completion of readout, which includes (i) request/acknowledge propagation along the arbitration tree and (ii) any queueing delay when multiple sources contend. Properly choosing \(T\) (the acknowledge quantum), the number of trees/outputs and the arbitration depth bound this service time and, therefore, the effective time-stamp uncertainty. This is fundamentally different from frame-based or matrix-freeze approaches that limit resolution to the frame interval regardless of instantaneous activity. With appropriately chosen event-driven parameters, sub-microsecond and often tens-of-nanoseconds, effective timing can be achieved.

Prior work on Address-Event Representation (AER) architectures analyzed timing mainly from the \emph{arbitration/bus} perspective rather than through a single centralized queue. In particular, \cite{Boahen_2000} quantified the latency and temporal dispersion on the arbiter-tree/bus links under burst-ensemble traffic and derived bandwidth conditions to preserve spike-timing precision, while \cite{ZamarreoRamos2011ModularAS} treated AER as a traffic/queueing problem at the encoder/arbiter chain to estimate latency, queueing delay, and occupancy. In short, AER timing has been studied, but \emph{system-level queueing theory for event-driven architecture has not been widely or successfully applied}, and existing analyzes emphasize \emph{distributed contention} rather than a monolithic buffer.

Building on that background, we previously experimented with canonical queueing-model abstractions known from Queueing Theory but found they \emph{fail at medium-to-high load}. Specifically, \(M/G/1/K\)\footnote{Kendall's notation~\cite{Bhat2015}} (finite buffer with Poisson input)~\cite{Keilson1993} places losses at a \emph{central} queue and folds path delays into the server time, and \(M/G/1//N\) (Engset; finite population, also denoted as \(M/G/1/K/N\), where $K = \infty$)~\cite{Takine_Takagi_Hasegawa_1993} suppresses arrivals when many sources are busy. Neither captures the \emph{per-source one-slot blocking} that dominates in pixelated structures, such as radiation detectors, where each pixel can hold only one pending request and the server completes at most one job per acknowledge period \(T\) whenever work is present. Consequently, these models agree with measurements at low rates but \emph{diverge near saturation}, mispredicting pile-up, utilization, and sojourn time.

Motivated by these shortcomings, this work introduces and validates a \emph{tractable, physics-faithful model} for an event-driven readout system based on the EDWARD architecture. We model the readout core as an \(M/D/1\) server with deterministic quantum \(T\), and describe the input via \emph{per-source one-slot gating} that thins the admitted Poisson arrivals without invoking a central buffer. From this construction, we derive closed-form expressions for (i) mean sojourn time as a function of \(T\), arbitration depth/delays, and aggregate load; (ii) probability of pile-up (loss) per-source; (iii) utilization and throughput; and (iv) design guidance for selecting the number of trees/outputs and serialization speed to avoid saturation. Analytical results and simulations jointly explain why other models fail at higher occupancy and quantify the timing-resolution gains achievable with properly dimensioned event-driven designs like EDWARD.

\section{Working Principle of Event-Driven Readout with Arbiter Tree}
\label{sec:principle}

\subsection{Asynchronous request/acknowledge handshake}
\label{subsec:async-req-ack}

\begin{figure}[!t]
	\centering
	\includegraphics[width=\linewidth]{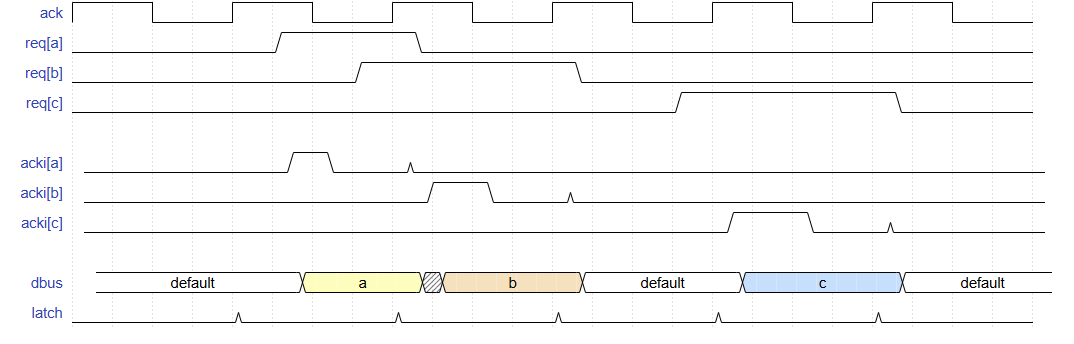}
	\caption{Asynchronous request/acknowledge handshake in the EDWARD architecture. Three pixels (\emph{a} -- \emph{c}) issue requests, and arbitration gates the global acknowledge \texttt{ack} to produce per-winner signals \texttt{acki[k]}. Only the granted pixel observes the two same-polarity edges required to start and complete a transfer. The data bus (\texttt{dbus}) carries each pixel's payload in non-overlapping windows (\emph{a} -- \emph{c}). Unassigned intervals (\emph{default}) correspond to idle bus states, while hatched regions denote short bus turnaround times. The \texttt{latch} marks sampling instants used by the serializer -- one completion is available per server quantum \(T\).}
	\label{fig:async-handshake}
\end{figure}

In the EDWARD readout architecture, each pixel asserts a \emph{request} (\texttt{req}[k]) when a validated event (e.g., a particle hit) occurs. An arbitration tree grants exclusive access to the shared readout bus to at most one requester at a time (see Figure~\ref{fig:async-handshake}). Bus access is transacted by a two-edge handshake using a global acknowledge (\texttt{ack}) that is distributed back to the pixels through the arbitration tree. The tree behaves like a clock-gating network and delivers a per-winner gated acknowledge, \texttt{acki}[k]. The first active edge of \texttt{acki}[k] begins the transfer by enabling the drivers from the selected pixel that drive data onto the data bus. Then the second active edge, of the same polarity, completes the transfer, causes the pixel to self-reset, and latches the data in the serializer located in the periphery. While requests are asynchronous with respect to the acknowledge timing, the downstream server (bus + serializer) operates in fixed quanta of duration \(T\), completing at most one transfer per period whenever work (a request) is present.

\subsection{Cell-level arbitration with Seitz mutexes}
\label{subsec:cell}

\begin{figure}[!ht]
	\centering
	\includegraphics[width=0.5\linewidth]{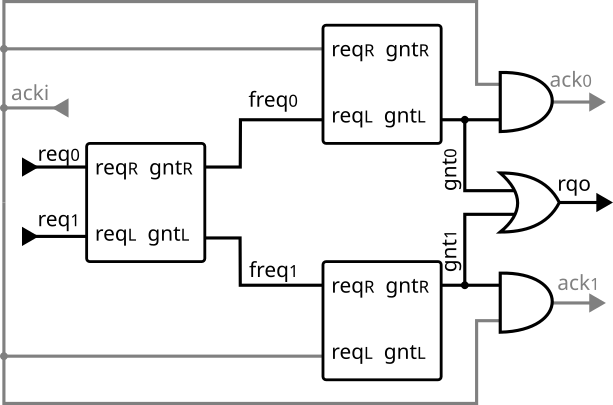}
	\caption{Binary arbiter cell based on Seitz mutexes. Left: child requests (\texttt{req0}, \texttt{req1}) are resolved into one-hot outputs (\texttt{freq0}, \texttt{freq1}). Right: per-branch acknowledge interlocks qualify local requests with the \texttt{ack} to prevent in-cell races or glitches. OR/AND networks form the upward propagated request and the gated acknowledges toward the children~\cite{Gorni_2022}.}
	\label{fig:arb-cell}
\end{figure}

Each binary arbiter \emph{cell} is built from Seitz RS-latch based mutual-exclusion elements (mutexes)~\cite{Seitz_1980} arranged to avoid glitches and races when \texttt{ack} is in flight. Functionally, the cell contains \emph{three} arbiters, as shown in Figure~\ref{fig:arb-cell}:

\begin{enumerate}
	\item A request arbiter that resolves the two child requests (\texttt{req0}, \texttt{req1}) into one-hot local requests (\texttt{freq0}, \texttt{freq1}).
	\item Two acknowledge interlock arbiters (one per child) that qualify the local request with the parent acknowledge to ensure that the cell does not change state while \texttt{ack} is asserted inside the cell.
	\item The combination guarantees a \emph{two-step} update: (i) decide the winner on requests; (ii) release and retime updates only after \texttt{ack} has been observed low again at the cell, thereby preventing hazards and race conditions on the upward request and downward acknowledge paths.
\end{enumerate}

This organization enforces local First-Come, First-Served (FCFS) arbitration: once a child’s request state is captured, the opposite child is blocked until the handshake completes. When the winning pixel resets on the second \texttt{acki} edge, its request is cleared, allowing a pending child request to propagate but only after \texttt{ack} is withdrawn by the parent.

\subsection{Tree composition, path clearing, and fairness}
\label{subsec:tree}

\begin{figure}[!ht]
	\centering
	\includegraphics[width=0.6\linewidth]{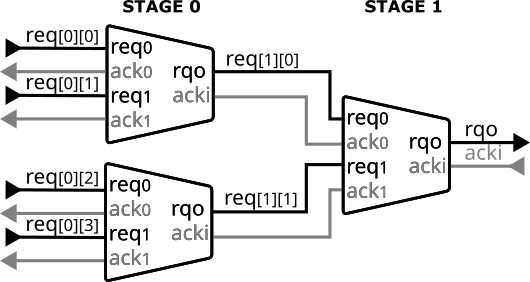}
	\caption{Two-stage arbitration tree. Stage~0 cells arbitrate leaf pairs, and Stage~1 arbitrates between their winners, forming the complete request/acknowledge hierarchy.}
	\label{fig:arb-tree}
\end{figure}

Cells are organized into a binary tree whose leaves connect to pixels and whose root connects to the acknowledge generator. The two-stage arbitration tree is shown in Figure~\ref{fig:arb-tree}. A decision propagates upward, stage by stage, toward the root, after which the \texttt{ack} signal is back-propagated downward along the single selected path using the arbitration mechanism described earlier.

Path clearing is triggered by request resets. Upon completion, the \emph{leaf} withdraws its request (self-resetting on the second \texttt{acki} edge). Each cell along the winning path then clears its output request, if only momentarily, in the case where a second leaf request is active -- before passing control to the next stage. As a result, if a sibling branch is actively requesting, it is immediately revealed to the higher level and becomes eligible to win arbitration. This request withdrawal propagates upward like a domino toward the root, which in turn removes the gating condition for \texttt{ack} along that path (i.e., \texttt{ack} is \emph{withdrawn} from the path that has just completed).

In aggregate, this yields:

\begin{itemize}
	\item No global FCFS guarantee (decisions are made on a \emph{per-cell} basis).
	\item Practical starvation avoidance via request-driven path clearing: a branch that just won tends to defer to its sibling until that sibling is serviced, approximating round-robin among active sub-trees without global state.
\end{itemize}

\subsection{Example waveforms}
\label{subsec:waveform}

\begin{figure}[!ht]
	\centering
	\includegraphics[width=\linewidth]{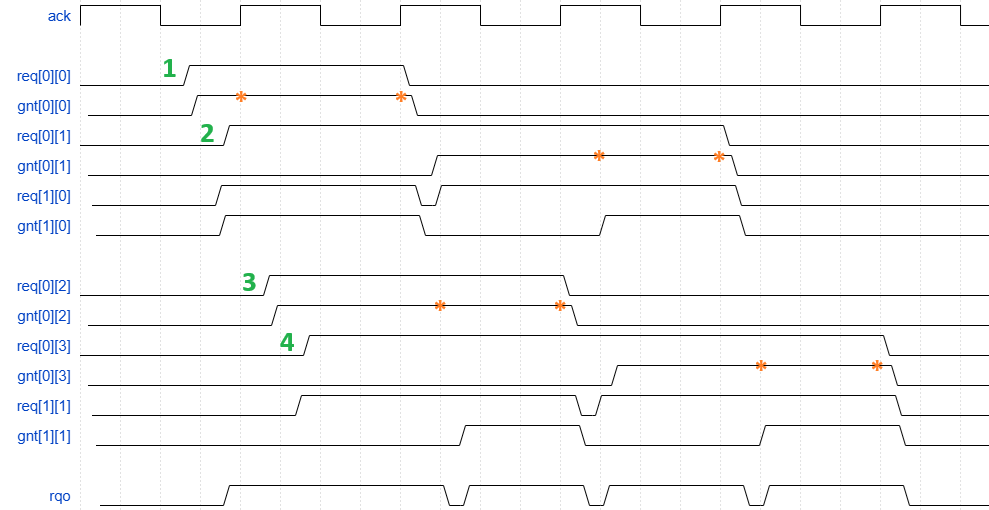}
	\caption{Timing of a four-leaf arbiter tree under overlapping requests. Labels \textbf{1}--\textbf{4} denote \emph{request arrivals}. The actual service sequence is \textbf{1}--\textbf{3}--\textbf{2}--\textbf{4}. Orange asterisks mark pixel-level actions triggered by the acknowledge edges.}
	\label{fig:fcfs-desc}
\end{figure}

Figure~\ref{fig:fcfs-desc} illustrates the timing of a four-leaf tree under overlapping requests. The green numbers \textbf{1}--\textbf{4} mark successive \emph{request arrivals}; due to local FCFS and the path clearing mechanism, the \emph{actual service order} is \textbf{1} \(\rightarrow\) \textbf{3} \(\rightarrow\) \textbf{2} \(\rightarrow\) \textbf{4}. Orange asterisks mark the instants when pixel actions are triggered (latch on first \texttt{acki} edge and self-reset on the second). The global \texttt{ack} is periodic with quantum \(T\), so at most one completion occurs per \(T\). The acknowledge path is omitted on the waveforms for clarity but in reality its effect is visible through the gated \texttt{acki[k]} pulses that only appear on the winner’s branch.

It is important to note that there is \emph{no dead time between consecutive readouts} -- within the same \texttt{ack} period, the system completes the readout from one pixel and immediately begins servicing another, provided that a new request is pending.

\section{Analytical Model}
\label{sec:model}

This section develops a tractable and experimentally validated analytical model for event-driven readout with arbiter trees. The fundamental observation is that the \emph{server side} of the system (the shared bus and serializer) completes exactly one request per acknowledge period~\(T\) whenever any work is present. The service process is therefore \emph{deterministic} at the granularity of~\(T\). Conversely, event losses (pileup) occur \emph{locally at each source}, since every source can buffer at most one pending request. These properties naturally lead to an \(M/D/1\) queue fed by a \emph{thinned} Poisson stream that accounts for the per-source one-slot gating.

\subsection{Modeling assumptions and notation}

We introduce the following symbols:

\begin{itemize}
	\item \(N\): number of independent sources (pixels or channels);
	\item \(\lambda\): per-source Poisson mean arrival rate (events/s);
	\item \(T\): acknowledge period, corresponding to the deterministic server quantum;
	\item \(L=\lceil \log_2 N\rceil\): number of arbitration stages in the binary tree;
	\item \(t_1,t_2\): mean per-stage forward (req-up) and backward (ack-down) delays;
	\item \(\tau_g\): fixed logic overhead;
	\item \(\tau_0 = L(t_1+t_2) + \tau_g\): path-dependent propagation delay from leaf to root and back;
	\item \(U \sim \mathrm{Unif}[T/2,3T/2]\): alignment jitter arising from the asynchronous phase between an event and the acknowledge clock edges;
	\item \(S\): server service time.
\end{itemize}

The total event-to-completion (sojourn) time observed at a source is therefore
\begin{equation}
	S_t = \tau_0 + U + W_q,
	\label{eq:St-decomp}
\end{equation}
where \(W_q\) is the queueing delay at the root (in multiples of~\(T\)), and the server service time itself is deterministic, \(S\equiv T\).

\subsection{Timing decomposition with tree delays}

Propagation delays in the arbitration tree contribute a fixed overhead,
\begin{equation}
	\tau_0 = L(t_1+t_2) + \tau_g,
	\label{eq:tau0}
\end{equation}
while asynchronous timing relative to the global acknowledge clock introduces a uniformly distributed jitter \(U \sim \mathrm{Unif}[T/2,3T/2]\). For an isolated request (no contention), the event-to-completion time simplifies to 
\begin{equation}
	S_t = \tau_0 + U,
	\label{eq:single}
\end{equation}
yielding the bounds \(S_{t,\min} = \tau_0 + T/2\), \(S_{t,\max} = \tau_0 + 3T/2\), and the mean value \(\mathbb{E}[S_t] = \tau_0 + T\). These single-pixel measurements provide direct experimental calibration of \(\tau_0\).

\subsection{Per-source one-slot gating and admitted rate}

Each source can store only one pending request -- new hits that occur while a request is awaiting acknowledge are irreversibly lost. According to the \emph{Poisson Arrivals See Time Averages} (PASTA)~\cite{Wolff_1982} principle, the fraction of arrivals that find the source busy equals the fraction of time the source is busy. This leads to
\begin{equation}
	P_{loss} = \frac{\lambda\,\mathbb{E}[S_t]}{1+\lambda\,\mathbb{E}[S_t]},
	\qquad
	1-P_{loss} = \frac{1}{1+\lambda\,\mathbb{E}[S_t]},
	\label{eq:pr}
\end{equation}
where \(P_{loss}\) is the per-source pileup probability. Summing over all sources gives the total admitted rate at the root:
\begin{equation}
	\boxed{\;
	\Lambda = N\lambda(1-P_{loss}) = \frac{N\lambda}{1+\lambda\,\mathbb{E}[S_t]}\; ,}
	\label{eq:Lambda}
\end{equation}
which can be treated as effectively having the Poisson nature for large \(N\) due to the weak cross-correlation between sources.

\subsection{Root queue: M/D/1 core}

Given the admitted rate \(\Lambda\) and the deterministic service time~\(T\), the root behaves as an \(M/D/1\) queue. Let \(\rho = \Lambda T\) denote the utilization. For a general \(M/G/1\) system under FCFS scheduling, the mean waiting time is
\begin{equation}
	\mathbb{E}[W_q] = \frac{\Lambda \mathbb{E}[S^2]}{2(1-\rho)}.
	\label{eq:Wq_gen}
\end{equation}
With deterministic service \(S\equiv T\),
\begin{equation}
	\mathbb{E}[W_q] = \frac{\Lambda T^2}{2(1-\Lambda T)} = \frac{\rho\,T}{2(1-\rho)}.
	\label{eq:Wq}
\end{equation}

Combining \eqref{eq:St-decomp} and \eqref{eq:Wq} and noting \(\mathbb{E}[U]=T\), we obtain
\begin{equation}
	\boxed{\;
	\mathbb{E}[S_t] = \tau_0 + T + \frac{\Lambda T^2}{2(1-\Lambda T)}\; .}
	\label{eq:Est}
\end{equation}

Equations \eqref{eq:Lambda}--\eqref{eq:Est} form a \emph{self-consistent system} linking the mean sojourn time \(\mathbb{E}[S_t]\), the admitted rate \(\Lambda\), the pileup probability \(P_{loss}\), and the utilization~\(\rho\).

\subsection{Closed-form solution}

The coupled equations above can be solved algebraically to obtain an explicit, \emph{closed-form} expression for \(\mathbb{E}[S_t]\) without iterative numerical methods.

Let \(a = N\lambda\), \(b = T\), and \(\tau = \tau_0\). Substituting \eqref{eq:Lambda} into \eqref{eq:Est} leads to a quadratic equation in a transformed variable \(u = 1 + \lambda\mathbb{E}[S_t] - ab\):
\begin{equation}
	u^2 - \big(1 - ab + \lambda(\tau+b)\big)u - \frac{\lambda a b^2}{2} = 0.
	\label{eq:uquad}
\end{equation}
The physically valid (positive) root is
\begin{equation}
	u = \frac{C + \sqrt{C^2 + 2\lambda a b^2}}{2},
	\qquad
	C = 1 - ab + \lambda(\tau + b).
	\label{eq:uroot}
\end{equation}
Finally,
\begin{equation}
	\boxed{\;
	\mathbb{E}[S_t] = \frac{u - 1 + ab}{\lambda}\; ,}
	\label{eq:Est-closed}
\end{equation}
and the remaining quantities follow directly:
\begin{equation}
	\boxed{\;
	P_{loss} = \frac{\lambda\,\mathbb{E}[S_t]}{1+\lambda\,\mathbb{E}[S_t]},\quad
	\Lambda = \frac{N\lambda}{1+\lambda\,\mathbb{E}[S_t]},\quad
	\rho = \Lambda T\; .}
	\label{eq:derived}
\end{equation}

In the next section, we will prove that this closed-form, self-consistent model accurately predicts experimental measurements across both light and heavy traffic regimes, providing a quantitative bridge between asynchronous arbitration and queueing-theoretic performance metrics.

\subsection{Stability and asymptotics}

The \(M/D/1\) queue is stable provided that the effective service rate exceeds the admitted arrival rate, \(\rho=\Lambda T<1\). Substituting \eqref{eq:Lambda} gives
\begin{equation}
	N\lambda T < 1+\lambda\,\mathbb{E}[S_t].
\end{equation}

This inequality is inherently satisfied by the self-consistent solution of Eqs.~\eqref{eq:Lambda}--\eqref{eq:Est}, since \(\mathbb{E}[S_t]\) increases with load in a way that limits \(\Lambda\) such that \(\rho<1\).

For \textbf{light traffic, i.e., \(\lambda\to 0\)} (when arrivals are sparse), the system behaves as a collection of independent sources. The total admitted rate is \(\Lambda \approx N\lambda\), the utilization grows linearly with the rate, and the mean sojourn time converges to the single-source timing:
\begin{equation}
	S_{t,\min}=\tau_0+\tfrac{T}{2},\qquad
	S_{t,\max}=\tau_0+\tfrac{3T}{2},\qquad
	\overline{S}_t=\tau_0+T .
	\label{eq:calib}
\end{equation}
These relations can directly calibrate the effective propagation delay~\(\tau_0\) from one-pixel measurements.

For \textbf{heavy traffic i.e., \(\lambda\to\infty\)} as the input rate increases, the throughput saturates at one completion per acknowledge period: \(\Lambda\!\to\!1/T\), \(\rho\!\to\!1\), and nearly all sources hold active requests (\(P_{loss}\!\to\!1\)). In this limit, the system behaves like a round-robin scheduler that services each of the \(N\) sources once per cycle. Consequently, the mean sojourn time per source does not diverge as in a conventional queue but remains bounded by the time needed to serve all sources once,
\begin{equation}
	\boxed{\;\mathbb{E}[S_t]_{\max} \approx N\,T + \tau_0\;,}
\end{equation}
which matches the observed saturation of \(S_t\) at high load. This finite upper bound reflects the fairness of the arbitration tree: each source is guaranteed service within approximately \(N\) acknowledge periods, avoiding unbounded queueing delays typical of centralized buffers. The bounded behavior ensures predictable latency even at full occupancy, a key property for architectural scaling.

\subsection{Summary of computable outputs}

Given the system parameters \((N,\,\lambda,\,T)\), the fixed delay is obtained either from experimental calibration using~\eqref{eq:single} or~\eqref{eq:calib} or be estimated based on technology timing using~\eqref{eq:tau0}. Solving Eqs.~\eqref{eq:uquad}--\eqref{eq:Est-closed} yields the mean sojourn time~\(\mathbb{E}[S_t]\), from which the loss probability~\(P_{loss}\), the admitted rate~\(\Lambda\), and the utilization~\(\rho\) follow via~\eqref{eq:derived}. These closed-form relations enable rapid exploration of design trade-offs (tree size, serialization rate, acknowledge period) while reproducing the observed system behavior across a wide range of loads.

\section{Model validation}
\label{sec:validation}

The self-consistent \(M/D/1\) model with per-source admission control was validated against (i) post-layout simulations of an EDWARD-class prototype~\cite{Gorni_2025} and (ii) software-based discrete-event simulations of the asynchronous arbiter tree. For comparison, two classical abstractions were also included: the \(M/G/1/K\) model (finite central buffer with \(K{=}N\)) and the Engset \(M/G/1//N\) model (finite population). Additionally, an analytical \(M/G/1\) variant was evaluated assuming a uniform service-time distribution over \([\tau_0 + T/2,\,\tau_0 + 3T/2]\), while preserving the same admission rule as the \(M/D/1\) core.

\subsection{Model parameters}

\begin{figure}[t]
	\centering
	\includegraphics[width=0.5\linewidth]{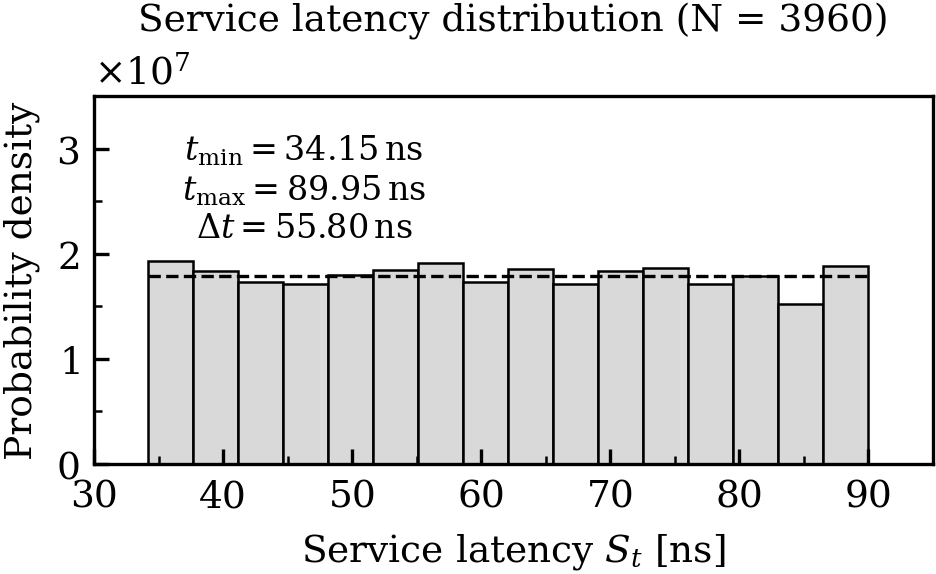}
	\caption{Distribution of the service latency \(S_t\) for a single isolated pixel. The histogram’s position and width provide estimates of the intrinsic request/acknowledge path delay \(\tau_0\) and timing jitter, independent of inter-pixel contention~\cite{Gorni_FEE2023}.}
	\label{fig:svc-onepixel}
\end{figure}

To ensure comparability with post-layout simulation data, the analytical and simulated models were parameterized according to the prototype specifications. For the sake of reference, the prototype comprises \(N = 1,024\) pixels and operates with an external serialization clock of 250\,MHz, which is internally divided by 14 to generate the acknowledge and data-latch signals in the serializer. This configuration yields a deterministic service quantum of \(T = 56~\mathrm{ns}\). The prototype allows pixel-level activation control and, by enabling a single pixel, the intrinsic service latency \(\tau_0\) can be extracted from relatively fast post-layout simulations. Figure~\ref{fig:svc-onepixel} presents the distribution of the service latency \(S_t\) for an isolated pixel. From these results, the intrinsic request/acknowledge path delay was estimated as \(\tau_0 = 6.05~\mathrm{ns}\).

\subsection{Post-layout reference data}
To obtain reference data for model validation, three full-scale post-layout simulations of the EDWARD prototype were performed under distinct input-rate conditions~\cite{Gorni_FEE2023}: low, medium, and high. Each scenario corresponds to a different per-pixel event-generation rate \(\lambda\) and an effective total arrival rate \(\Lambda^{*}\), calculated as a sum of all pixel rates in the absence of contention.

In the \emph{low-rate} scenario, the per-pixel rate was \(\lambda = 948.2~\mathrm{s^{-1}}\), giving an aggregate \(\Lambda^{*} = 970.9~\mathrm{ks^{-1}}\). The corresponding mean inter-arrival time \(1/\Lambda^{*} \approx 1.02~\mathrm{\mu s}\) was much larger than the service quantum \(T = 56~\mathrm{ns}\), i.e., \(1/\Lambda^{*} \gg T\). Under such sparse traffic, almost all pixel requests were served immediately by the next available acknowledge pulse, with minimal queueing effects.

In the \emph{medium-rate} scenario, \(\lambda = 15{,}169.2~\mathrm{s^{-1}}\) and \(\Lambda^{*} = 15.53~\mathrm{Ms^{-1}}\), yielding \(1/\Lambda^{*} \approx 64~\mathrm{ns}\), which becomes comparable to \(T\). In this intermediate regime, contention between pixels starts to play a significant role. The average service delay increases as multiple requests compete for acknowledge, resulting in a broader latency distribution with occasional quantized delays corresponding to integer multiples of \(T\).

In the \emph{high-rate} scenario, the pixel-level rate reached \(\lambda = 239.98~\mathrm{ks^{-1}}\), giving a total \(\Lambda^{*} = 24.57~\mathrm{Ms^{-1}}\). Here, \(1/\Lambda^{*} \approx 40.7~\mathrm{ns} \ll T\), implying that new events arrive faster than acknowledges can be issued. In this saturation regime, nearly every pixel experiences waiting time, and the mean latency asymptotically approaches the full-matrix readout time -- as it was a frame-by-frame acquisition process where each pixel must wait for all others to be serviced.

\begin{figure}[t]
	\centering

\begin{subfigure}[t]{\textwidth}
	\centering
	\begin{subfigure}[t]{0.32\textwidth}
		\includegraphics[width=\linewidth]{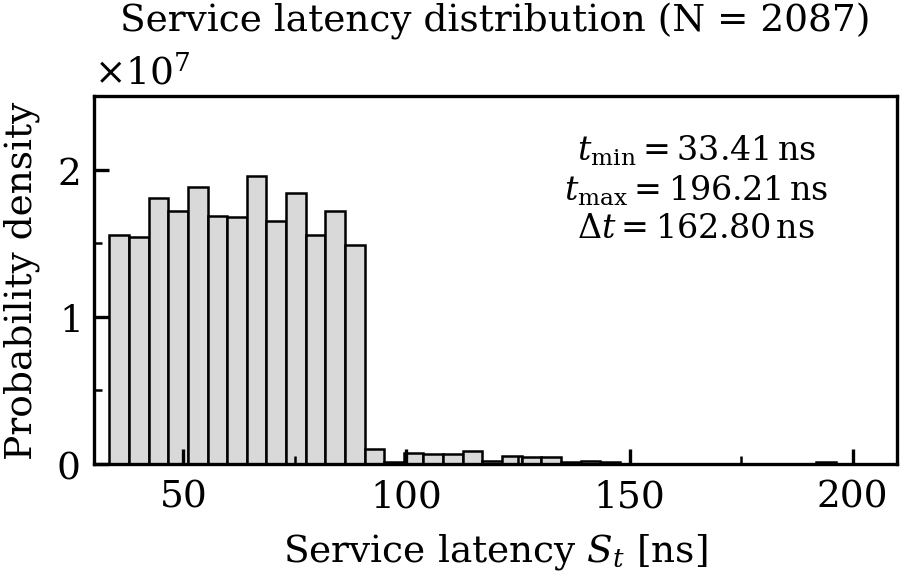}
		\caption{Low-rate.}
		\label{fig:svc-low}
	\end{subfigure}\hfill
	\begin{subfigure}[t]{0.32\textwidth}
		\includegraphics[width=\linewidth]{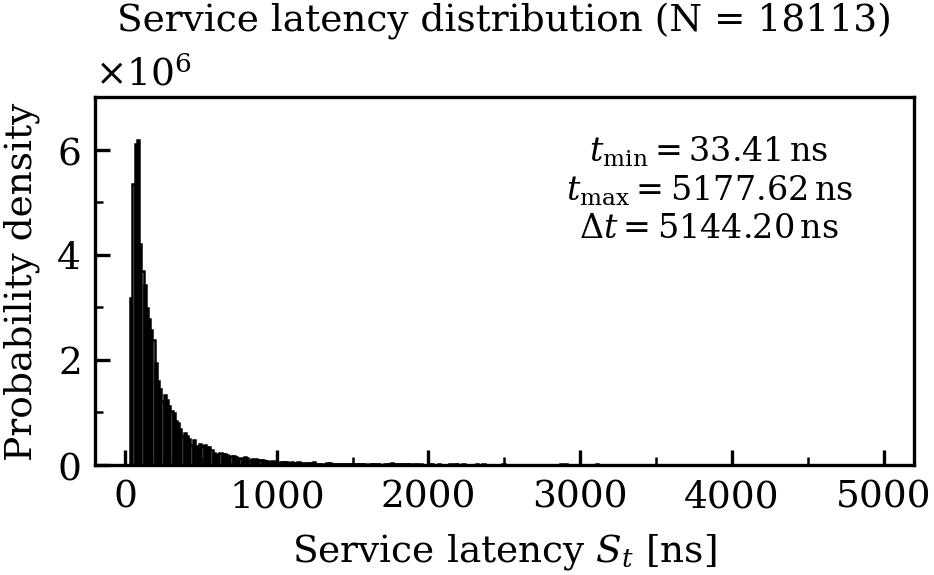}
		\caption{Medium-rate.}
		\label{fig:svc-med}
	\end{subfigure}\hfill
	\begin{subfigure}[t]{0.32\textwidth}
		\includegraphics[width=\linewidth]{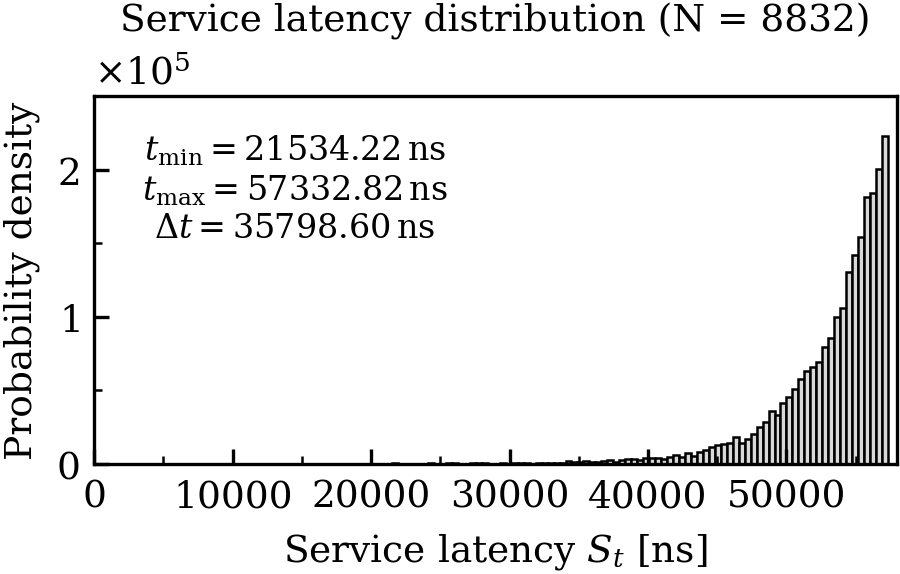}
		\caption{High-rate.}
		\label{fig:svc-high}
	\end{subfigure}
	\caption*{(A) Global \emph{Sojourn-time} distributions \(S_t\) shown from sparse (a) to contention-limited (c) operation.}
	\end{subfigure}


	\begin{subfigure}[t]{\textwidth}
	\centering
	\begin{subfigure}[t]{0.32\textwidth}
		\includegraphics[width=\linewidth]{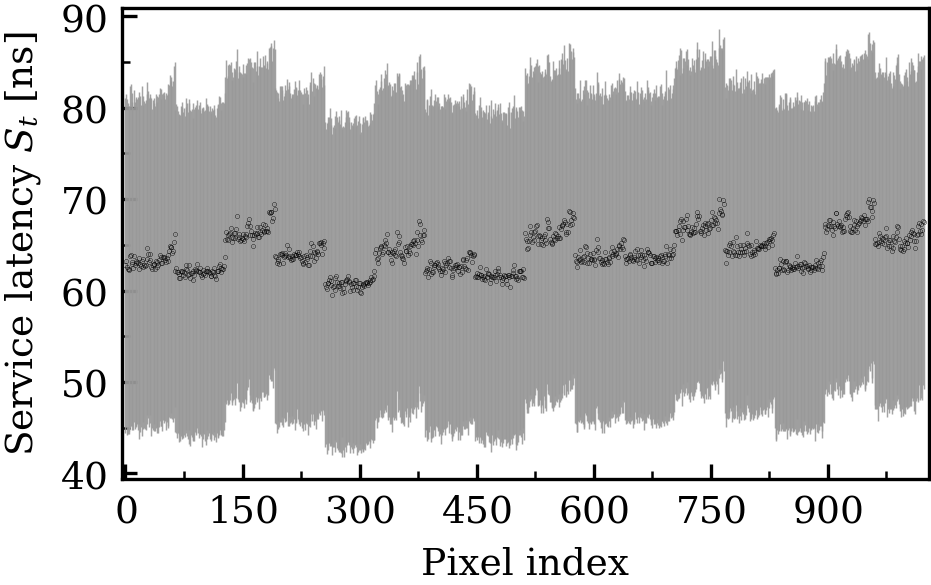}
		\caption{Low-rate.}
		\label{fig:latency-low}
	\end{subfigure}\hfill
	\begin{subfigure}[t]{0.32\textwidth}
		\includegraphics[width=\linewidth]{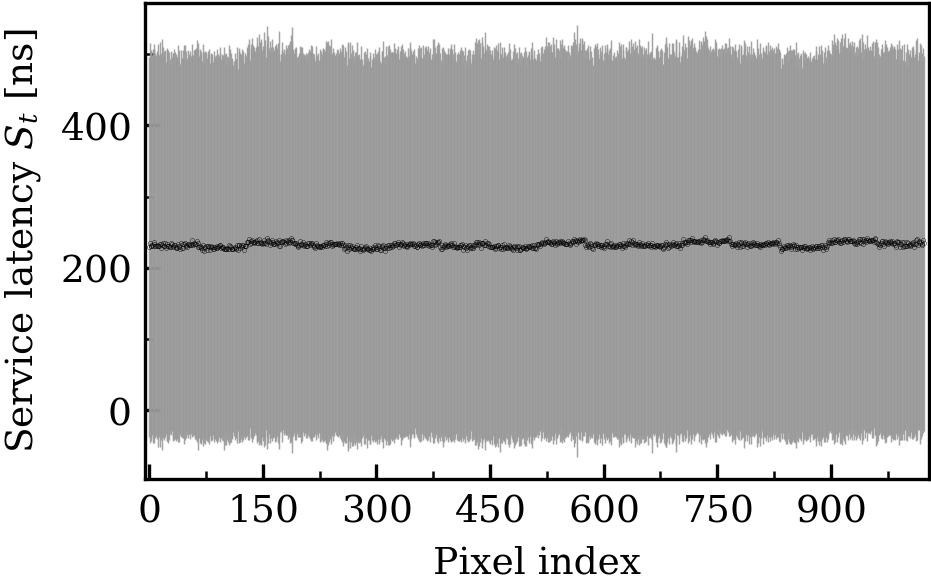}
		\caption{Medium-rate.}
		\label{fig:latency-med}
	\end{subfigure}\hfill
	\begin{subfigure}[t]{0.32\textwidth}
		\includegraphics[width=\linewidth]{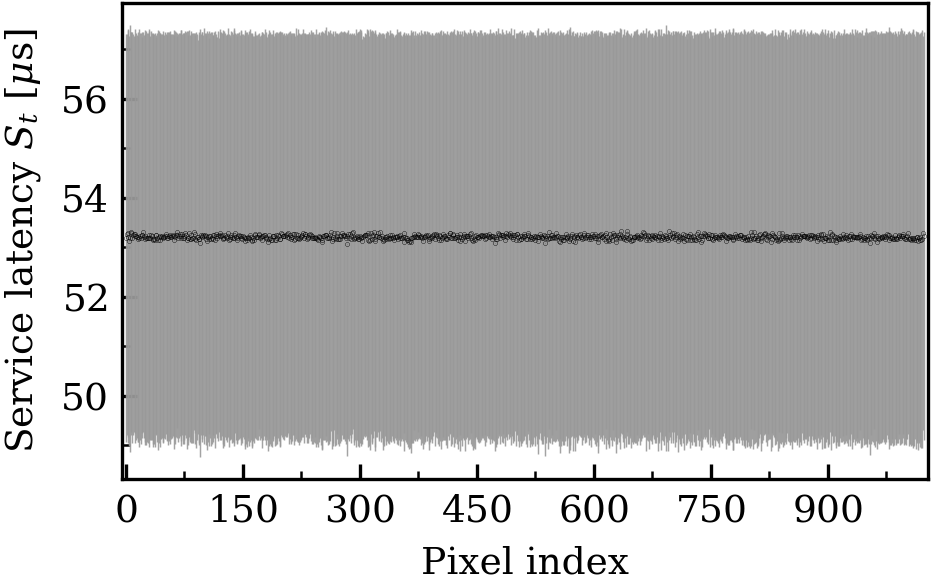}
		\caption{High-rate.}
		\label{fig:latency-high}
	\end{subfigure}
	\caption*{(B) Per-pixel mean \emph{service latency} \(S_t\) maps reveal uniform handshake delays, with small layout-dependent shifts. Each black marker denotes mean value and thin gray whiskers indicate \(\pm1\sigma\) over repeated transients.}
	\end{subfigure}


	\begin{subfigure}[t]{\textwidth}
	\centering
	\begin{subfigure}[t]{0.32\textwidth}
		\includegraphics[width=\linewidth]{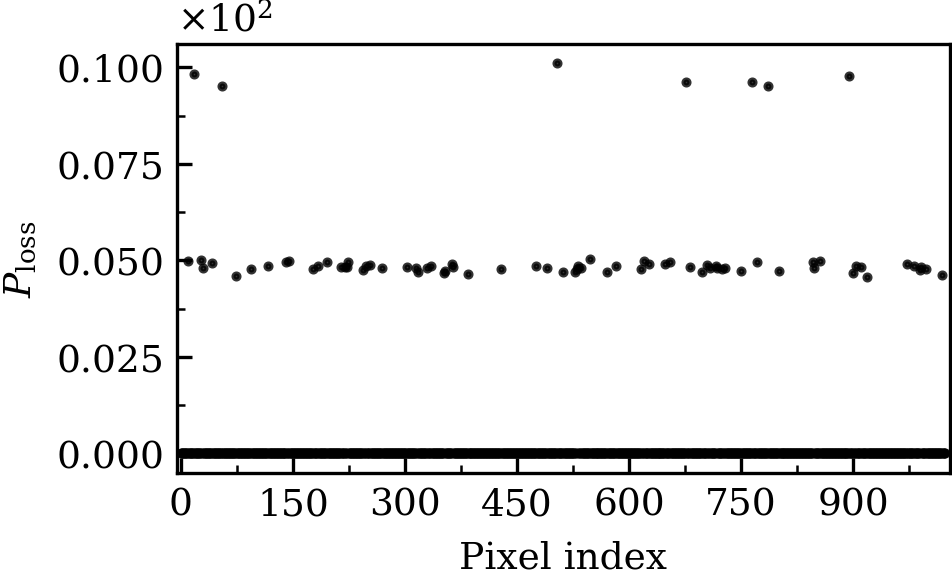}
		\caption{Low-rate.}
		\label{fig:pileup-low}
	\end{subfigure}\hfill
	\begin{subfigure}[t]{0.32\textwidth}
		\includegraphics[width=\linewidth]{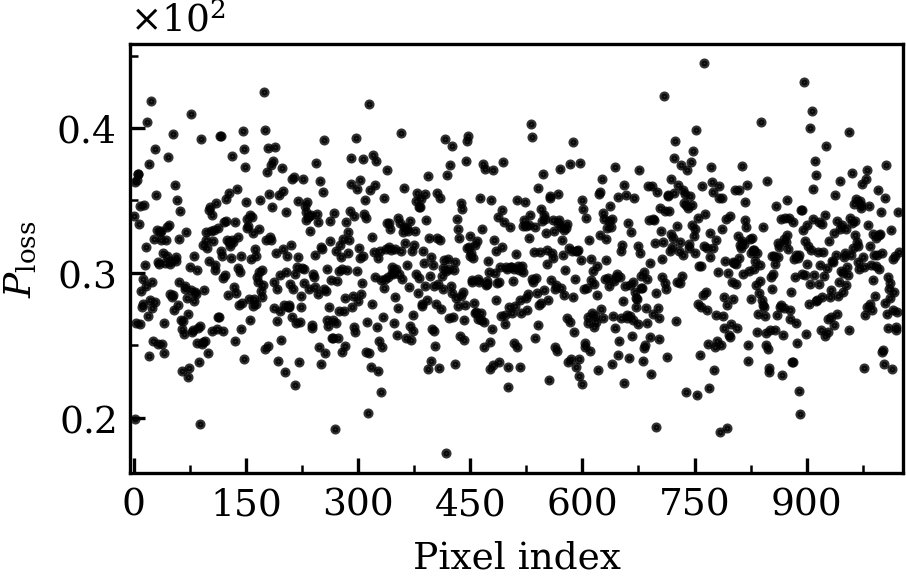}
		\caption{Medium-rate.}
		\label{fig:pileup-med}
	\end{subfigure}\hfill
	\begin{subfigure}[t]{0.32\textwidth}
		\includegraphics[width=\linewidth]{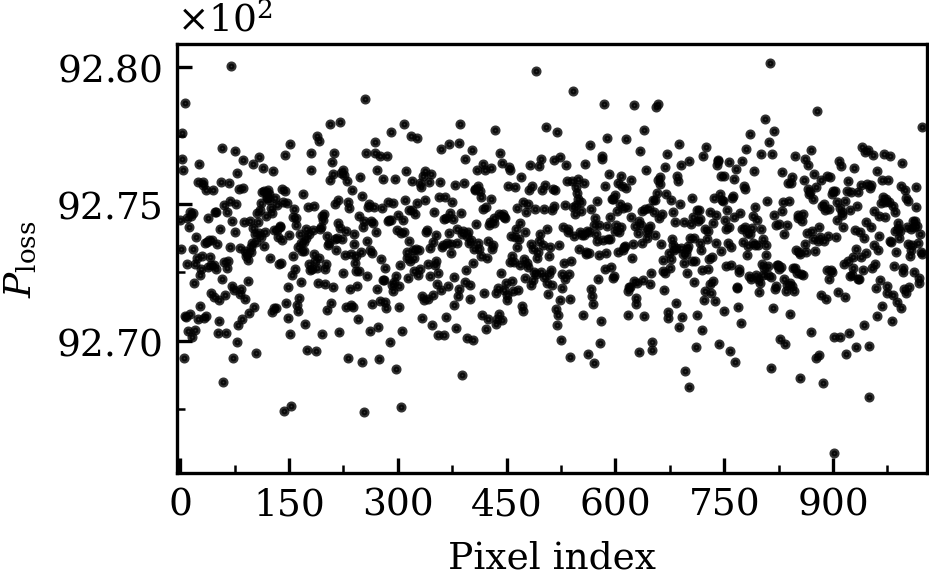}
		\caption{High-rate.}
		\label{fig:pileup-high}
	\end{subfigure}
	\caption*{(C) Per-pixel \emph{pile-up probability} \(P_{\mathrm{loss}}\) confirms globally uniform admission behavior shown from sparse (a) to contention-limited (c) operation.}
	\end{subfigure}

	\caption{Comprehensive post-layout validation under three input-rate regimes~\cite{Gorni_FEE2023}.}
	\label{fig:validation-triptych}
\end{figure}

The recorded distributions of the service (sojourn) times for these three operating regimes are presented in Figure~\ref{fig:validation-triptych}(A). The evolution of the distribution, from a narrow, nearly uniform shape at low \(\lambda\) to a right-skewed, quantized profile at high \(\lambda\), clearly reflects the transition from sparse, asynchronous operation to contention-limited throughput.

Complementary insight is provided by the per-pixel maps shown in Figure~\ref{fig:validation-triptych}(B--C). Panel (B) presents the spatial distribution of the mean service latency \(S_t\) across all 1,024 pixels for each input-rate regime. At low rates, the map is essentially uniform, confirming that all handshake paths within the arbiter tree contribute nearly identical propagation delays. Minor spatial variations (\(<\!\!1~\mathrm{ns}\)) can be attributed to systematic routing differences and transistor-level parasitics in the physical layout. As the input rate increases, these latency maps remain globally flat, indicating that contention is evenly distributed among the pixels and that the arbitration network operates without introducing geometric bias or priority artifacts.

Panel (C) illustrates the corresponding per-pixel pile-up probability \(P_{\mathrm{loss}}\), defined as the ratio of lost (not handled) requests to the total number of the generated events. As expected, the loss probability grows monotonically with \(\lambda\): it is practically zero in the low-rate regime, reaches the sub-percent level at medium rates, and approaches unity in the high-rate case. The near-uniform spatial distribution of \(P_{\mathrm{loss}}\) further confirms that the admission process depends solely on global load and not on pixel position or electrical distance within the matrix.

\subsection{Modeling results}

\begin{figure}[htb]
	\centering
	\begin{subfigure}[b]{0.48\linewidth}
		\centering
		\includegraphics[width=\linewidth]{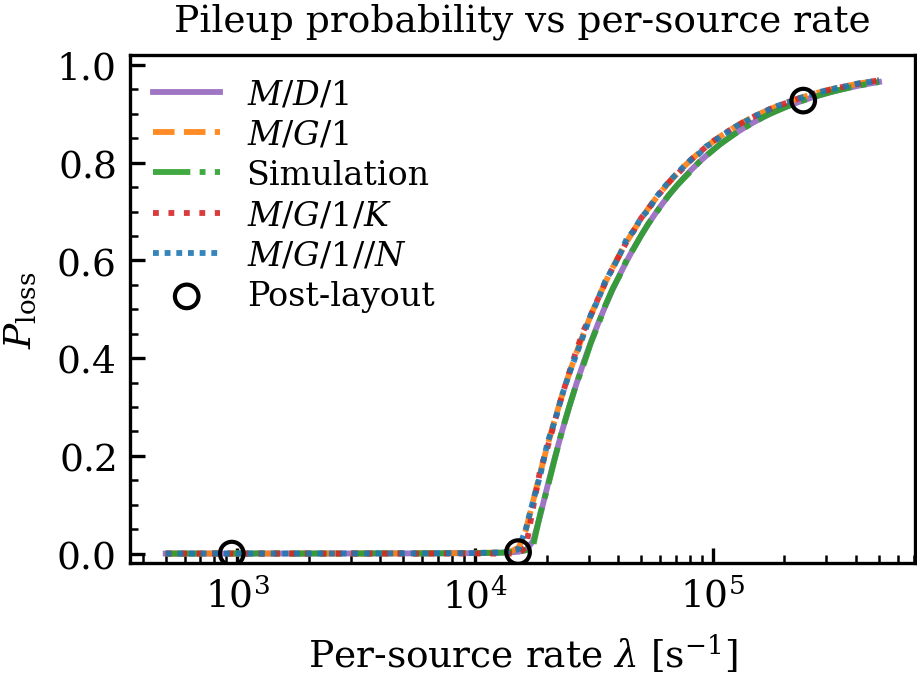}
		\caption{Probability of pile-up \(P_{\mathrm{loss}}\) as a function of per-source rate \(\lambda\). \\}
		\label{fig:overlay_pileup_all}
	\end{subfigure}
	\hfill
	\begin{subfigure}[b]{0.48\linewidth}
		\centering
		\includegraphics[width=\linewidth]{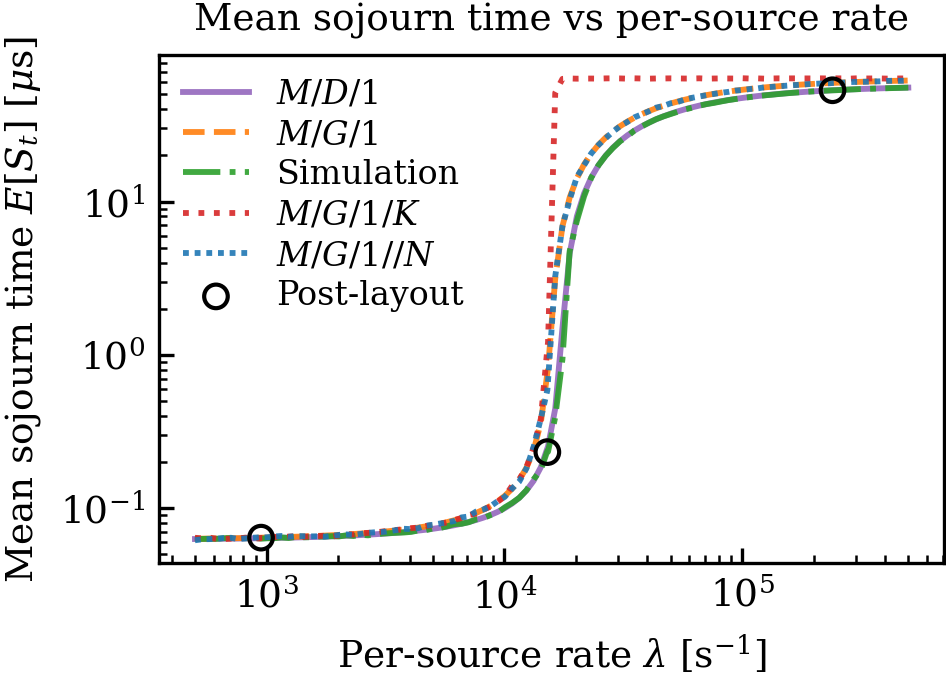}
		\caption{Mean sojourn time \(\mathbb{E}[S_t]\) as a function of per-source rate \(\lambda\). The deterministic-service model reproduces the growth up to saturation.}
		\label{fig:overlay_ESt_all}
	\end{subfigure}


	\begin{subfigure}[b]{0.48\linewidth}
		\centering
		\includegraphics[width=\linewidth]{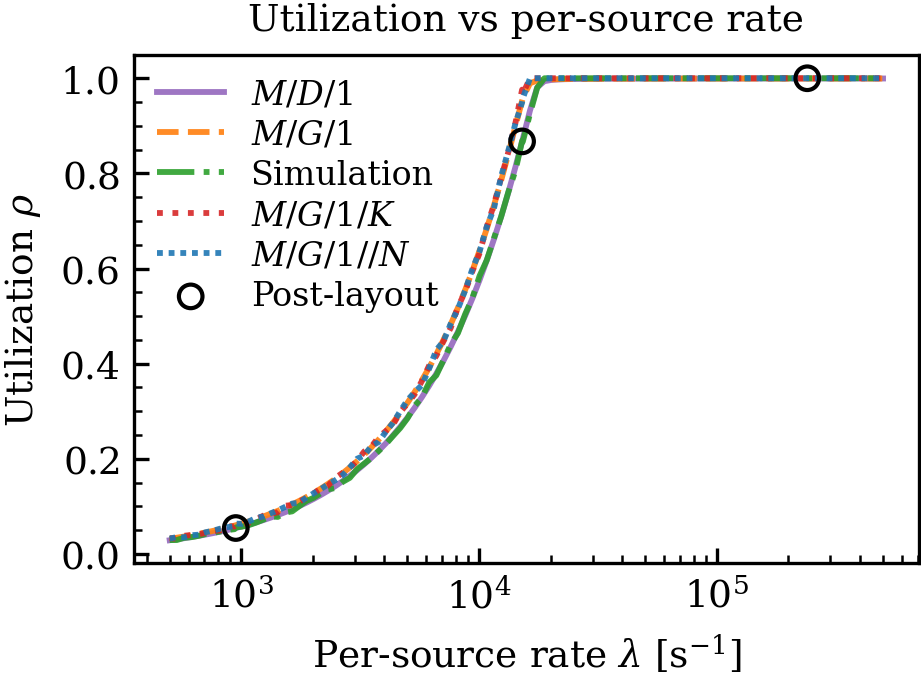}
		\caption{Utilization \(\rho\) (fraction of active ack quanta) as a function of per-source rate \(\lambda\). All models grow nearly linearly at low load.}
		\label{fig:overlay_rho_all}
	\end{subfigure}
	\hfill
	\begin{subfigure}[b]{0.48\linewidth}
		\centering
		\includegraphics[width=\linewidth]{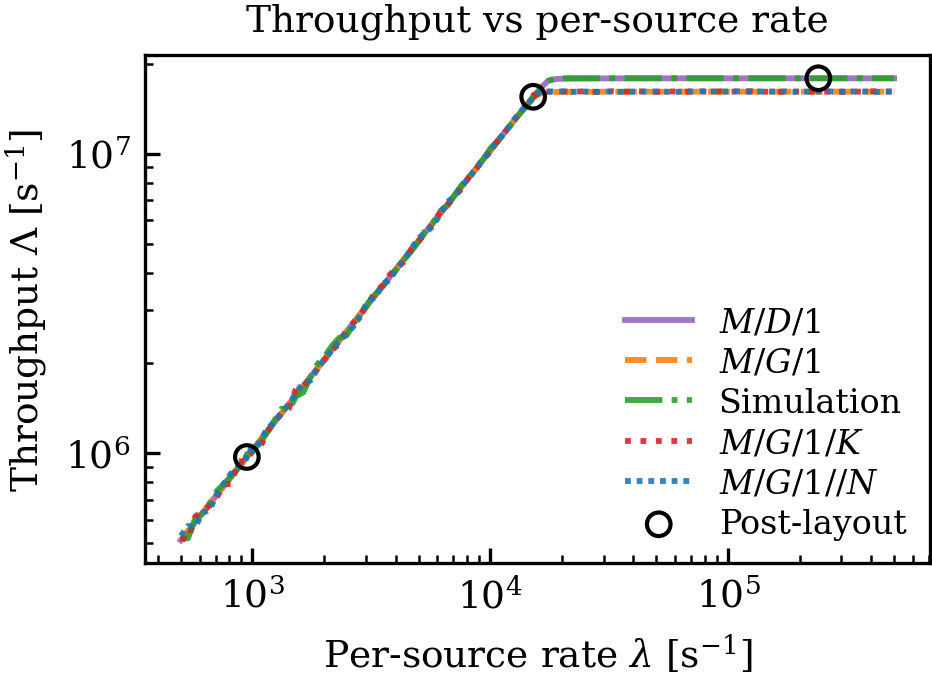}
		\caption{Aggregate throughput \(\Lambda\) as a function of per-source rate \(\lambda\). Simulation and \(M/D/1\) approach the theoretical limit \(1/T\) as the system backlogs.}
		\label{fig:overlay_throughput_all}
	\end{subfigure}

	\caption{Comparison of analytical models, simulation, and measurements as a function of per-source rate \(\lambda\).}
	\label{fig:overlay_all}
\end{figure}

Table~\ref{tab:validation-summary} reports mean sojourn time \(\mathbb{E}[S_t]\), per-source pile-up probability \(P_{\mathrm{loss}}\), utilization \(\rho\), and aggregate throughput \(\Lambda\) at three representative per-source rates. Figures~\ref{fig:overlay_all}(a--d) summarize the full-range trends.

\begin{table}[t]
	\centering
	\caption{Comparison of post-layout, simulation, and analytical models at three representative per-source rates (\(N=1,024\), \(T=56~\mathrm{ns}\), \(\tau_0=6.05~\mathrm{ns}\)).}
	\label{tab:validation-summary}
	\vspace{0.6ex}
	\begin{tabular}{r l r r r r}
	\toprule
	\(\lambda\) [\(\mathrm{s^{-1}}\)] & Model & \(\mathbb{E}[S_t]\) [s] & \(P_{\mathrm{loss}}\) & \(\rho\) & \(\Lambda\) [\(\mathrm{s^{-1}}\)]\\
	\midrule
	948
		& Post-layout			& \(6.420\times10^{-8}\)	& \(0.000\%\)	& \(5.437\%\)	& \(9.710\times10^{5}\) \\
		& \textbf{\(M/D/1\)}	& \(6.366\times10^{-8}\)	& \(0.006\%\)	& \(5.437\%\)	& \(9.709\times10^{5}\) \\
		& \(M/G/1\)				& \(6.417\times10^{-8}\)	& \(0.006\%\)	& \(6.024\%\)	& \(9.709\times10^{5}\) \\
		& Simulation			& \(6.403\times10^{-8}\)	& \(0.000\%\)	& \(5.347\%\)	& \(9.547\times10^{5}\) \\
		& \(M/G/1/K\)			& \(6.403\times10^{-8}\)	& \(0.000\%\)	& \(5.856\%\)	& \(9.427\times10^{5}\) \\
		& \(M/G/1//N\)			& \(6.485\times10^{-8}\)	& \(0.006\%\)	& \(6.077\%\)	& \(9.713\times10^{5}\) \\
	\midrule
	15{,}169
		& Post-layout			& \(2.320\times10^{-7}\)	& \(0.300\%\)	& \(86.725\%\)	& \(1.549\times10^{7}\) \\
		& \textbf{\(M/D/1\)}	& \(2.440\times10^{-7}\)	& \(0.369\%\)	& \(86.665\%\)	& \(1.548\times10^{7}\) \\
		& \(M/G/1\)				& \(7.367\times10^{-7}\)	& \(1.105\%\)	& \(95.319\%\)	& \(1.536\times10^{7}\) \\
		& Simulation			& \(2.451\times10^{-7}\)	& \(0.389\%\)	& \(86.889\%\)	& \(1.552\times10^{7}\) \\
		& \(M/G/1/K\)			& \(8.436\times10^{-7}\)	& \(0.000\%\)	& \(96.076\%\)	& \(1.549\times10^{7}\) \\
		& \(M/G/1//N\)			& \(6.323\times10^{-7}\)	& \(0.956\%\)	& \(95.713\%\)	& \(1.548\times10^{7}\) \\
	\midrule
	240{,}000
		& Post-layout			& \(5.320\times10^{-5}\)	& \(92.700\%\)	& \(100.000\%\)	& \(1.786\times10^{7}\) \\
		& \textbf{\(M/D/1\)}	& \(5.321\times10^{-5}\)	& \(92.737\%\)	& \(99.947\%\)	& \(1.785\times10^{7}\) \\
		& \(M/G/1\)				& \(5.941\times10^{-5}\)	& \(93.445\%\)	& \(99.944\%\)	& \(1.611\times10^{7}\) \\
		& Simulation			& \(5.316\times10^{-5}\)	& \(92.732\%\)	& \(99.999\%\)	& \(1.786\times10^{7}\) \\
		& \(M/G/1/K\)			& \(6.346\times10^{-5}\)	& \(93.422\%\)	& \(100.000\%\)	& \(1.613\times10^{7}\) \\
		& \(M/G/1//N\)			& \(5.912\times10^{-5}\)	& \(93.440\%\)	& \(100.000\%\)	& \(1.618\times10^{7}\) \\
	\bottomrule
	\end{tabular}
\end{table}

At \textbf{low rate} all models agree that the system is lightly loaded, with \(\rho\approx 5\)--\(6\%\) and essentially with zero loss. The prediction \(\mathbf{M/D/1}\) matches the post-layout \(\Lambda\) and \(\rho\) to within \(<\!0.01\%\) absolute and \(\mathbb{E}[S_t]\) within \(<\!1\%\). The software \textbf{simulation} is statistically coincident. Small systematic offsets in \(\Lambda\) for \(\mathbf{M/G/1/K}\) and \(\mathbf{M/G/1//N}\) reflect their different admission mechanisms (central blocking vs.\ finite-population suppression), not the source-level pile-up.

At \textbf{medium rate} contention becomes appreciable: \(\rho\) rises to \(\sim 0.87\) in the post-layout data and \(\mathbf{M/D/1}\), with modest loss (\(P_{\mathrm{loss}}\approx 0.3\%\)--\(0.4\%\)). Here \(\mathbf{M/D/1}\) slightly \emph{overestimates} delay: \(2.44\times 10^{-7}\) vs.\ \(2.32\times 10^{-7}\) s, by $\sim$\ 5\% and \(P_{\mathrm{loss}}\) by \(\sim 0.07\) percentage points (pp), while still tracking \(\Lambda\) and \(\rho\) closely. This may be the result of limited simulation time of the post-layout run, especially when the software \textbf{Simulation} shows a comparable deviation. In contrast, the uniform-service \(\mathbf{M/G/1}\) inflates both \(\rho\) and delay (\(\mathbb{E}[S_t]\) is larger by a factor \(\sim 3\)), because it folds \(\tau_0\) and alignment into the server time. \(\mathbf{M/G/1/K}\) shows negligible blocking at this \(K{=}N\) (loss \(\approx 0\%\)) but still lengthens \(\mathbb{E}[S_t]\) via central buffering. Engset \(\mathbf{M/G/1//N}\) behaves similarly to \(\mathbf{M/G/1}\) in delay and utilization owing to finite-population suppression effects.

At \textbf{high rate} the system saturates as expected: \(\Lambda \to 1/T \approx 1.786\times 10^{7}\,\mathrm{s^{-1}}\), \(\rho \to 1\), and \(P_{\mathrm{loss}}\to 92.7\%\), with \(\mathbb{E}[S_t]\approx 53.2~\mu\mathrm{s}\). \(\mathbf{M/D/1}\) and \textbf{Simulation} coincide with the post-layout data within the reported spread. Models that treat \(\tau_0\) as server occupancy (uniform-service \(\mathbf{M/G/1}\), \(\mathbf{M/G/1/K}\), and Engset \(\mathbf{M/G/1//N}\)) underestimate throughput by \(\sim 9\)--\(10\%\) and exhibit longer mean sojourn times, being consistent with their inflated effective service times and, for \(M/G/1/K\), central blocking.

Across all regimes, the dominant behaviors are dictated by the architectural facts established in Section~\ref{sec:model}: (i) a \emph{deterministic} server at the root (one completion per \(T\) when backlogged); (ii) \emph{per-source one-slot} admission, which places loss at the sources rather than in a central queue; and (iii) \(\tau_0\) and phase alignment adds to sojourn time but does \emph{not} consume root service. The \(\mathbf{M/D/1}\) abstraction preserves all three, hence it remains quantitatively predictive from light to heavy traffic. By contrast, uniform-service \(\mathbf{M/G/1}\), \(\mathbf{M/G/1/K}\), and Engset \(\mathbf{M/G/1//N}\) each violate at least one of these architectural constraints, leading to the observed overestimates in delay and shortfalls in saturation throughput.

\section{Using the Model During System Design}
\label{sec:design}

The analytical framework developed in Section~\ref{sec:model} can be employed to predict system performance under different architectural configurations. Because the relations are algebraic and self-consistent, they allow rapid evaluation of latency, loss probability, utilization, and throughput without time-domain simulation. The model is suitable for:

\begin{itemize}
	\item preliminary dimensioning of readout architectures under specified limits on latency, loss, and throughput;
	\item analysis of the impact of acknowledge period \(T\), arbitration depth, and intrinsic delay \(\tau_0\);
	\item selection of the number and size of independent subsystems (tiles) for a given I/O and power budget.
\end{itemize}

In large matrices, the readout system may be divided into \(c\) independent tiles, each containing \(N_j\) sources and its own arbiter and serializer. Substituting \((N,\lambda,T,\tau_0)\!\rightarrow\!(N_j,\lambda_j,T_j,\tau_{0,j})\) in the analytical relations yields tile-specific quantities \(\mathbb{E}[S_t]_j\), \(P_{\mathrm{loss},j}\), and \(\rho_j\). The total throughput is the sum of tile throughputs, \(\Lambda_{\mathrm{tot}}=\sum_j\Lambda_j\). Reducing the tile size decreases the arbitration depth \(L_j=\lceil\log_2 N_j\rceil\) and consequently the propagation delay \(\tau_{0,j}\), improving latency and reducing local pile-up at constant \(T_j\). A typical use of the model involves:

\begin{enumerate}
	\item defining the array size \(N\), expected per-pixel rate \(\lambda\), and acknowledge period \(T\);
	\item estimating per-tile path delays \(\tau_{0,j}\) from layout or measurement;
	\item computing per-tile metrics \(\mathbb{E}[S_t]_j\), \(P_{\mathrm{loss},j}\), and \(\rho_j\) using the closed-form relations;
	\item verifying that all tiles satisfy target limits \(P_{\mathrm{loss},j}\le P_{\mathrm{loss}}^{\max}\), \(\mathbb{E}[S_t]_j\le \mathbb{E}[S_t]^{\max}\), and \(\rho_j<1\);
	\item iterating on tile size \(N_j\) or acknowledge period \(T_j\) if constraints are not met.
\end{enumerate}

These steps allow for an early architectural exploration before a detailed simulation.

\section{Summary and Conclusions}
\label{sec:summary}

This work formulated a tractable analytical description of arbiter-tree, event-driven readout with a synchronous acknowledge quantum. The readout core was modeled as an \(M/D/1\) server (one completion per acknowledge period \(T\) when backlogged), while losses were placed at the sources via one-slot gating. The resulting self-consistent relations link admitted rate \(\Lambda\), mean sojourn time \(\mathbb{E}[S_t]\), utilization \(\rho\), and loss probability \(P_{\mathrm{loss}}\), with a closed form for \(\mathbb{E}[S_t]\) that separates fixed path delay \(\tau_0\) and queueing.

Validation against an EDWARD-class prototype and software simulations showed agreement across low, medium, and high load, including saturation at \(1/T\). In contrast, abstractions that centralize blocking or fold \(\tau_0\) into service time deviated at moderate and high occupancy. The model provided quantitative guidance for design: tile-level partitioning reduces arbitration depth and \(\tau_{0}\), lowers latency and pile-up at fixed \(T\), and scales aggregate throughput additively across tiles.

Future extensions include percentile metrics for \(S_t\), explicit burst models beyond Poisson, non-uniform rate maps with tiling optimization, and co-design of serializer timing and arbitration for energy-latency trade-offs. The proposed formulation bridges physical implementation and high-level performance modeling, offering a practical analytical tool for future event-driven readout designs.


\bibliographystyle{JHEP}
\bibliography{biblio.bib}

@ARTICLE{Boahen_2000,
  author={Boahen, K.A.},
  journal={IEEE Transactions on Circuits and Systems II: Analog and Digital Signal Processing}, 
  title={Point-to-point connectivity between neuromorphic chips using address events}, 
  year={2000},
  volume={47},
  number={5},
  pages={416-434},
  doi={10.1109/82.842110}}

@inproceedings{ZamarreoRamos2011ModularAS,
  title={Modular and scalable implementation of AER neuromorphic systems},
  author={Carlos Zamarre{\~n}o-Ramos},
  year={2011},
  url={https://api.semanticscholar.org/CorpusID:193527387}
}

@Article{He2023,
author={He, Rui
and Niu, Xiao-Yang
and Wang, Yi
and Liang, Hong-Wei
and Liu, Hong-Bang
and Tian, Ye
and Zhang, Hong-Lin
and Zou, Chao-Jie
and Liu, Zhi-Yi
and Zhang, Yun-Long
and Yang, Hai-Bo
and Huang, Ju
and Wang, Hong-Kai
and Han, Wei-Jia
and others},
title={Advances in nuclear detection and readout techniques},
journal={Nuclear Science and Techniques},
year={2023},
month={Dec},
day={21},
volume={34},
number={12},
pages={205},
issn={2210-3147},
doi={10.1007/s41365-023-01359-0}}

@article{Schayck_2022,
    author = {van Schayck, J Paul and Zhang, Yue and Knoops, Kèvin and Peters, Peter J and Ravelli, Raimond B G},
    title = {Integration of an Event-driven Timepix3 Hybrid Pixel Detector into a Cryo-EM Workflow},
    journal = {Microscopy and Microanalysis},
    volume = {29},
    number = {1},
    pages = {352-363},
    year = {2022},
    month = {12},
    issn = {1431-9276},
    doi = {10.1093/micmic/ozac009},
}

@ARTICLE{Yang_2019,
  author={Yang, Chenfei and Feng, Changqing and Liu, Jun and Teng, Yao and Liu, Shubin and An, Qi and Sun, Xiangming and Yang, Ping},
  journal={IEEE Transactions on Nuclear Science}, 
  title={A Prototype Readout System for the ALPIDE Pixel Sensor}, 
  year={2019},
  volume={66},
  number={7},
  pages={1088-1094},
  doi={10.1109/TNS.2019.2913335}}

@article{Gorni_2022,
doi = {10.1088/1748-0221/17/04/C04027},
url = {https://doi.org/10.1088/1748-0221/17/04/C04027},
year = {2022},
month = {apr},
publisher = {IOP Publishing},
volume = {17},
number = {04},
pages = {C04027},
author = {Gorni, D.S. and Deptuch, G.W. and Miryala, S. and Siddons, D.P. and Kuczewski, A. and Rumaiz, A.K. and Carini, G.A.},
title = {Event driven readout architecture with non-priority arbitration for radiation detectors},
journal = {Journal of Instrumentation}
}

@INPROCEEDINGS{Gorni_prime,
  author={Gorni, Dominik and Deptuch, Grzegorz and Miryala, Sandeep},
  booktitle={2022 17th Conference on Ph.D Research in Microelectronics and Electronics (PRIME)}, 
  title={Investigation of Timing Properties for an Event Driven with Access and Reset Decoder Readout Architecture for a Pixel Array}, 
  year={2022},
  volume={},
  number={},
  pages={113-116},
  doi={10.1109/PRIME55000.2022.9816805}}

@Article{Keilson1993,
author={Keilson, J.
and Servi, L. D.},
title={The M/G/1/K blocking formula and its generalizations to state-dependent vacation systems and priority systems},
journal={Queueing Systems},
year={1993},
month={Mar},
day={01},
volume={14},
number={1},
pages={111-123},
issn={1572-9443},
doi={10.1007/BF01153529},
url={https://doi.org/10.1007/BF01153529}
}

@article{Takine_Takagi_Hasegawa_1993, title={Analysis of an M/G/1/K/N queue}, volume={30}, DOI={10.2307/3214853}, number={2}, journal={Journal of Applied Probability}, author={Takine, T. and Takagi, H. and Hasegawa, T.}, year={1993}, pages={446–454}}

@book{Bhat2015,
  author    = {U. Narayan Bhat},
  title     = {An Introduction to Queueing Theory: Modeling and Analysis in Applications},
  series    = {Statistics for Industry and Technology},
  edition   = {2},
  year      = {2015},
  publisher = {Birkhäuser},
  address   = {Boston, MA},
  doi       = {10.1007/978-0-8176-8421-1},
  isbn      = {978-0-8176-8420-4}
}

@Article{Seitz_1980,
  author  = {Charles L. Seitz},
  journal = {Lambda},
  title   = {Ideas about arbiters},
  year    = {1980},
  number  = {1},
  pages   = {10--14},
}

@article{Wolff_1982,
author = {Wolff, Ronald W.},
title = {Poisson Arrivals See Time Averages},
journal = {Operations Research},
volume = {30},
number = {2},
pages = {223-231},
year = {1982},
doi = {10.1287/opre.30.2.223}
}

@article{Gorni_2025,
doi = {10.1088/1748-0221/20/03/C03009},
url = {https://doi.org/10.1088/1748-0221/20/03/C03009},
year = {2025},
month = {mar},
publisher = {IOP Publishing},
volume = {20},
number = {03},
pages = {C03009},
author = {Gorni, D.S. and Deptuch, G.W. and Maj, P. and Mandal, S. and Pinaroli, G.},
title = {Event-driven readout development: testing of the EDWARD65P1 chip with integrated event generators},
journal = {Journal of Instrumentation}
}

@inproceedings{Gorni_FEE2023,
  author       = {Dominik S. Gorni and Gabriella Carini and Grzegorz W. Deptuch
                  and Anthony Kuczewski and Soumyajit Mandal and Giovanni Pinaroli
                  and Abdul Rumaiz and Peter Siddons and Nicholas St. John},
  title        = {Event-Driven, Arbitrated Protocols Implemented in Integrated Readout Circuits for Segmented Sensors},
  booktitle    = {XII Front-End Electronics Workshop},
  address      = {Torino, Italy},
  month        = {June},
  year         = {2023},
  pages        = {},
  note         = {Oral presentation},
  organization = {INFN Torino},
  url          = {https://agenda.infn.it/event/36206/contributions/202624/},
}

\end{document}